\documentclass[11pt]{article}

\def\be{\begin{equation}}
\def\ee{\end{equation}}
\def\ba{\begin{eqnarray}}
\def\ea{\end{eqnarray}}
\def\nn{\nonumber}
\def\lb{\label}
\def\bb{\bibitem}
\def\E{{\cal E}}
\def\m{\overline\mu}
\def\v{\overline{v}}
\def\s{\sqrt{9-m^2}}
\def\mm{m_{\rm max}}

\begin{document}
\begin{titlepage}

\date{17 June 2019}

\title{
\begin{flushright}\begin{small}    LAPTH-034/19
\end{small} \end{flushright} \vspace{1cm}
Balanced magnetized double black holes}

\author{G\'erard Cl\'ement\thanks{Email: gclement@lapth.cnrs.fr} \\ \\
{\small LAPTh, Universit\'e Savoie Mont Blanc, CNRS,} \\ {\small 9 chemin de Bellevue,
BP 110, F-74941 Annecy-le-Vieux cedex, France}}

\maketitle

\begin{abstract}
We report on a two-parameter family of exact asymptotically flat stationary solutions to
Einstein-Maxwell theory.
These solutions are free from ring singularities in a specific parameter range.
They represent systems of two non-rotating extreme black holes with opposite magnetic and NUT charges,
connected by a tensionless Dirac-Misner string. For a given ADM mass of the system, the distance
between the two black holes has a lower bound. We also discuss the limit in which the two
constituent black holes are far apart.
\end{abstract}
\end{titlepage}
\setcounter{page}{2}

\section{Introduction}

Complementary to the intense ongoing activity in the observation of binary black hole mergers,
analytical investigations of stationary double black hole systems can help to understand
the physics of slowly evolving double black hole systems. It is well known that, generically,
exact stationary double black hole solutions to Einstein-Maxwell theory exhibit a conical
singularity (strut or cosmic string) on the portion of the axis between the two black holes,
which accounts for the force necessary to balance their mutual attraction or repulsion.
Exceptions are the static Majumdar-Papapetrou
(MP) \cite{MP} linear superpositions of extreme Reissner-Nordstr\"{o}m black holes. Stationary
asymptotically flat Israel-Wilson-Perj\`es (IWP) \cite{IWP} superpositions of two extreme
Reissner-Nordstr\"{o}m-NUT black holes \cite{brill}, with equal masses and electric charges,
and opposite NUT (gravimagnetic)
and magnetic charges, are also strutless. These are usually considered to be unphysical
because of the presence of a metrical line singularity (Misner string) connecting the two NUT
sources \cite{HH}. However we have shown elsewhere \cite{GC15} that, contrary to cosmic strings,
Misner strings are transparent to geodesic motion. And, while they are surrounded by a region
containing closed timelike curves (CTCs), we have argued that these do not necessarily lead to
observable violations of causality \cite{GC15,GC18}.

Both MP and IWP linear superpositions of two black holes with given physical characteristics (masses
and charges) exist for an arbitrary separation between the two black holes. In a recent paper
\cite{dimagn}, we have investigated a class of four-parameter stationary asymptotically flat
solutions to the Einstein-Maxwell
equations previously constructed in \cite{manko00a}, and shown that these represent non-linear
superpositions of two rotating extreme black holes, their separation being intrincately related
to the four black hole charges. Possible ring singularities can be avoided by an appropriate
restriction of the parameter domain, though this was not addressed in the general case in \cite{dimagn}.
Interestingly, we found some one- and two-parameter subclasses in which the conical singularity
is also absent in the horizon co-rotating frame, and pointed out that this does necessarily mean that
the system is balanced. The purpose of the present paper is to address systematically these two questions:
What is the parameter domain in which the ring singularity is absent ? And, is there a parameter
subclass in which the strut singularity is unambiguously absent ?

In the next section, we briefly review the four-parameter class of solutions as presented in
\cite{dimagn}, and determine a necessary condition for the absence of ring singularity. In section 3
we find a two-parameter subclass of solutions with vanishing conical singularity, both in the
horizon co-rotating frame and in the global frame (that of the observer at spacelike infinity). The
properties of these solutions, including the large-distance limit, are discussed in section 4. We
close with a brief summary of the results.

\setcounter{equation}{0}
\section{Review of the four-parameter family of double black holes}
We first review briefly the four-parameter family of magnetostatic solutions to the
Einstein-Maxwell equations constructed in \cite{manko00a} and interpreted
in \cite{dimagn} as the fields of double black hole systems. We recall that
the Ernst potentials\footnote{Our definitions of the Ernst potentials follow
\cite{smarr}, equations (3.1)-(3.3).}
$\E$ and $\psi$ may be expressed in terms of
Kinnersley potentials $(U,V,W)$ according to
 \be
{\cal E} = (U-W)/(U+W), \qquad \psi = V/(U+W),
 \ee
In the parametrization of \cite{dimagn}, the Kinnersley potentials of these
solutions are given in prolate spheroidal coordinates by:
 \ba\lb{kin}
U &=& (x^2-\delta y^2)^2 - d^2 - \nu\lambda(1-y^4)
+ 2ixy[\nu(x^2-1)+\lambda(1-y^2)], \nn\\
V &=& \m\{-\nu x(1-y^2) + iy[(x^2-1)+\delta(1-y^2)]\}, \nn\\
W &=& mx[(x^2-1)+(b\nu+\delta)(1-y^2)]- \nn\\ && -
imy[b(x^2-1)+(b\delta-\lambda)(1-y^2)].
 \ea
The prolate spheroidal coordinates $x\ge1$, $y\in[-1,+1]$ are related to
the Weyl cylindrical coordinates $\rho$, $z$ by
 \be\lb{sphero}
\rho = \kappa\sqrt{(x^2-1)(1-y^2)}, \quad z = \kappa xy,
 \ee
where $\kappa$ is the fundamental length scale. The other three independent
parameters may be chosen to be the dimensionless $m$, $\nu$ and $\lambda$,
the other parameters appearing in (\ref{kin}) being related to these by
 \ba
&& \delta = 1 + \nu^2 - \frac{m^2}4, \quad d = 1-\delta, \quad
\m^2 = m^2b^2 + 4d\delta, \lb{paras}\\
&& b = \frac2{m^2}\left[\nu(d-\delta) - \lambda\right] \lb{bla}
 \ea
(the reality of $\m$ is ensured only in a sector of the three-space $(m,\nu,\lambda)$).

These parameters are simply connected to the three main asymptotic observables
(read off from the behavior of (\ref{kin}) for $x\to\infty$), the total mass $M$,
angular momentum $J$ and dipole magnetic moment $\mu$:
 \be\lb{obs}
M = \kappa m, \quad J = \kappa^2 ma, \quad \mu = \kappa^2\m \qquad (a=b+2\nu).
 \ee
We will assume in the following $m\ge0$.
The total electric charge and electric dipole moment vanish, while the total
quadrupole electric moment is $Q_2 = - \kappa^2\m\nu$.

\subsection{Absence of ring singularity}

The solutions (\ref{kin}) can present a ring singularity where the function $U+W$
vanishes, so that the Ernst potentials $\E$ and $\psi$ blow up, signalling a
strong curvature singularity. So we should first restrict the model parameters to
the sector in which this ring singularity is absent\footnote{In \cite{dimagn} this
question was addressed only for special parameter subspaces.}. The ring singularity
corresponds to a solution of the system
 \be\lb{ring}
{\rm Re}(U+W)(x,y) = 0, \qquad {\rm Im}(U+W)(x,y) = 0
 \ee
with $y^2<1$. The second equation (\ref{ring}) is trivially satisfied in the
equatorial plane $z=0$ ($y=0$). As ${\rm Re}(U+W)$ is positive for large $x$,
a necessary condition for the absence of ring singularity is therefore
 \ba\lb{noring}
(U+W)(1,0) &=& 1 - d^2 - \nu\lambda + m(b\nu + \delta)  \nn\\
           &=& \delta(1+d+m) + \nu(mb-\lambda) \ge 0 .
 \ea

The parameters $b$ and $\lambda$ are linearly related through (\ref{bla}), so that,
using (\ref{paras}), the above condition may be rexpressed as a bound for $mb$:
 \be\lb{noring1}
mb \ge \frac{4\nu^2d-(m+2)^2\delta}{2\nu(m+2)},
 \ee
where we have assumed, without loss of generality, $\nu > 0$ (for the special case
$\nu=0$, see below). Interestingly, this regularity condition also
ensures that the scaled magnetic moment $\m$ is real, and gives the lower bound:
 \be
\m^2 \ge \left(\frac{4\nu^2d+(m+2)^2\delta}{2\nu(m+2)}\right)^2.
 \ee
In the special case $\nu=0$, treated in \cite{dimagn}, the ring singularity is
absent for $m<2$, all values of $b$ being allowed, and the condition $m<2$ also
ensures the reality of $\m$.

\subsection{Horizon and string}

The three Kinnersley potentials (\ref{kin}) and the corresponding stationary
metric and electromagnetic field degenerate for $x=1$, $y=\pm1$ ($\rho= 0,
z = \pm\kappa$).
The transformation to Kodama-Hikida \cite{KoHi} coordinates $(X,Y)$,
 \be\lb{XY}
X = \sqrt{\frac{1-y^2}{x^2-1}}, \qquad Y = \frac{y}x,
 \ee
resolves these coordinate singularities into two degenerate horizons $Y=\pm1$,
co-rotating at the angular velocity
 \be\lb{OmH}
\Omega_H = \frac{2(\nu-\lambda)}{\kappa[(m+2d)^2 + (mb-2\nu)^2]}.
 \ee
These two symmetrical surfaces, viewed in the co-rotating frame, are topological
spheres, smooth except at the
endpoint $X\to\infty$ where they generically present a conical singularity
with deficit angle $2\pi(1-\alpha_H)$ (the value of the constant $\alpha_H$
is given in \cite{dimagn}). The sign of this deficit angle is
difficult to determine in the general case.

Another coordinate singularity is the segment $x=1$, $y^2<1$ ($\rho=0$,
$-\kappa<z<\kappa$). The metric near this segment is that of a spinning
cosmic string in a background curved spacetime, with deficit angle
$é\pi(1-\alpha_S)$ and spin $-\omega_S/4$,
 \be\lb{string}
ds^2 \simeq -F_S(y)(dt-\omega_Sd\varphi)^2 + G_S(y)\left[
\frac{dy^2}{1-y^2}+d\xi^2+\alpha_S^2 \xi^2 d\varphi^2\right],
 \ee
where $\xi^2 \equiv x^2-1$, and
 \be
\alpha_S = \frac1{\delta^2+\nu\lambda},
 \ee
is generically different from $\alpha_H$. However, after transforming to
the horizon co-rotating frame by setting $d\varphi=d\hat\varphi
+ \Omega_Hdt$, one obtains for the cosmic string the transformed parameters
 \be\lb{stringpara}
\hat{\alpha}_S = \frac{\alpha_S}{(1-\Omega_H\omega_S)}, \qquad \hat\omega_S = \frac{\omega_S}
{(1-\Omega_H\omega_S)},
 \ee
where, not surprisingly,
 \be\lb{alphaSH}
\hat{\alpha}_S = \alpha_H.
 \ee
We recall that the deficit angle of the string metric is proportional (by a factor $8\pi$)
to the string tension, measuring the force between the two black holes, while the
string spin corresponds to a gravimagnetic flow along the
Misner string connecting two opposite NUT sources (the horizons) with
opposite NUT charges $\pm N_H$, with $N_H = - \hat\omega_S/4$,
 \be\lb{N}
N_H = \frac{\kappa\alpha_H}8\left[4\delta^2(mb-\lambda-\nu)
+ \nu(m^2b^2-4\lambda\nu)+2\delta(m+2d)(mb-2\lambda)-\lambda(m+2d)^2\right].
 \ee

The two horizons also carry equal Komar masses $M_H$, angular momenta $J_H$, and
electric charges $Q_H$ (computed in \cite{dimagn}), and opposite magnetic charges $\pm P_H$.
Knowing these, one obtains straightforwardly the string Komar mass $M_S=M-2M_H$, angular momentum
$J_S=J-2J_H$ and electric charge $Q_S=-2Q_H$.

\setcounter{equation}{0}
\section{Static class of solutions with vanishing string tension}

Generically, the string tension can be positive or negative, so that there is a class of solutions
with vanishing string tension. This class is such that, in the global frame, $\alpha_S=1$, or
 \be\lb{alphas1}
\delta^2+\nu\lambda = 1.
 \ee
However, as discussed in \cite{dimagn}, it is not clear whether this condition is enough to ensure
that the two constituent black holes exert no force on each other, or in other words that the
solution is singularity-free. This is because, from (\ref{stringpara}) and (\ref{alphaSH}),
$\alpha_H$ is generically different from $\alpha_S$ so that, if the string tension vanishes in
the global frame, the two horizons will nevertheless exhibit conical singularities in the local
horizon co-rotating frame. There are two possible ways to ensure $\alpha_H=\alpha_S$: either
$\omega_S=0$, or $\Omega_H=0$. The first condition, $\omega_S=0$, which means also $N_H=-\hat\omega_S/4=0$
(vanishing NUT charge), is often considered to be a necessary regularity condition (the ``axis condition").
We have argued elsewhere \cite{GC15} that this condition should be lifted, on the grounds that
Misner strings are not genuine singularities: they are
transparent to geodesic motion, and although they are surrounded by closed timelike curves,
the spacetime does not admit closed timelike geodesics. In the present case, we strongly doubt
whether the equation $N _H=0$ together with the condition (\ref{alphas1}) can lead to solutions free
from a ring singularity.

The other possibility is a vanishing horizon angular velocity,
 \be\lb{Om0}
\lambda = \nu.
 \ee
This condition selects the ``static" class of solutions discussed in \cite{dimagn}.
The vanishing string tension condition (\ref{alphas1}) then reduces to
 \be
\delta^2 + \nu^2 = 1,
 \ee
which may be combined with the first equation (\ref{paras}) to yield a second degree equation for
$\nu^2$ in terms of $m^2$, which is solved by
 \be
\nu^2 = \frac12\left[\frac{m^2}2-3 \pm \sqrt{9-m^2}\right],
 \ee
provided $m\le3$. As shown in \cite{dimagn}, in the static case the reality of $\m$ implies
$\delta\ge0$, which selects the upper sign, and $m \le 2\sqrt2$. One then finds, for the other
solution parameters,
  \be
\delta = \frac12\left[-1 + \sqrt{9-m^2}\right],  \quad d = \frac12\left[3 - \sqrt{9-m^2}\right]
 \ee
(both positive), and
 \be
mb = - \frac{4\nu\delta}m,\quad \m^2 = \frac{32d\delta}{m^2}.
 \ee

We wish to further constrain the sole dimensionless parameter $m$ to ensure the absence of ring singularity.
We first observe that
 \be
{\rm Im}(U+W)(x,y) = \nu y\left[2x(x^2-y^2) + \frac{4\delta}m(x^2-1) + \left(\frac{4\delta^2}m+1\right)(1-y^2)\right].
 \ee
The bracket is positive definite, so that the only possible ring singularities are
in the equatorial plane $y=0$. They will be absent if the function $Z(x) \equiv (U+W)(x,0)$
remains positive in the whole range $x \in [1,\infty[$. Using the above relations between
the solution parameters, we obtain
 \be
Z(x) = x^4 + mx^3 - \frac{8d}{m}x - 2d.
 \ee
One finds that $Z(1) \ge 0$ in the range $m\le2.506$. $Z'(x)$ is then positive for
$x=1$ and is an increasing function of $x$, so that the absence of ring singularities is guaranteed
by the condition
 \be
m \le \mm \simeq 2.506.
 \ee
The explicit metric and electromagnetic field are
  \ba\lb{ansatz}
ds^2 &=& - \frac{f}\Sigma\left(dt-\frac{\kappa\Pi}f d\varphi\right)^2 + \nn\\
&+& \kappa^2\Sigma\left[(x^2-y^2)^{-3}\left(\frac{dx^2}{x^2-1} +
\frac{dy^2}{1-y^2}\right) + f^{-1}(x^2-1)(1-y^2)d\varphi^2\right], \nn\\
A &=& \frac{1}{\Sigma}[\v dt + \kappa\Theta d\varphi],
 \ea
where the various functions are given by
 \ba\lb{sol}
&f(x,y) &= [\zeta^2 + \nu^2(1-y^2)^2]^2 - 4\nu^2(x^2-1)(1-y^2)(x^2-y^2)^2, \nn\\
&\Sigma(x,y) &= \left\{\zeta(\zeta+mx+2d) + mb\nu x(1-y^2) - \nu^2(1-y^4)\right\}^2 + \nn\\
&& + y^2\left\{2\nu x(x^2-y^2) + m[-b\zeta + \nu(1-y^2)]\right\}^2, \nn\\
&\Pi(x,y) &= - (1-y^2)\left\{\nu(x^2-1)( x^2-y^2)
\left(4mx[\zeta+mx+2d-b\nu(1+y^2)] + \right.\right. \nn\\
&& \left. + 2(m^2b^2-4d\delta)y^2\right)
+ [\zeta^2+\nu^2(1-y^2)^2]\cdot \nn\\
&&\cdot \left.\left(2mb(x+m)\zeta + \nu[-m(2x+m) +
\frac12(m^2b^2-4d\delta)](1-y^2)\right)\right\}, \nn\\
&\v(x,y) &= \m\left\{-\nu x(1-y^2)\left(\zeta(\zeta+mx+2d) + mb\nu x(1-y^2) - \nu^2(1-y^4)\right)\right. +\nn\\
&& \left. + y^2\zeta\left(2\nu x[x^2-y^2] + m[-b\zeta + \nu(1-y^2)]\right)\right\} \nn\\
&\Theta(x,y) &= \frac{\m(1-y^2)}2\left\{\left[\zeta(\zeta+mx+2d) + mb\nu x(1-y^2)
- \nu^2(1-y^4)\right]\cdot \right. \nn\\
&& \cdot \left[(2x+m)(\zeta + m^2) + 2m(x^2-d-2\nu^2)) - mb\nu(1+y^2)\right] + \nn\\
&& + 2y^2\left[2\nu x(x^2-y^2) + m(-b\zeta + \nu(1-y^2))\right] \cdot \nn\\
&& \left. \cdot \left[mb(x+m)-\nu(x^2-1)\right]\right\}.
\ea
In the preceding, we have put
 \be
\zeta \equiv x^2-1+\delta(1-y^2).
 \ee

\setcounter{equation}{0}
\section{Properties}

Although the horizons are non-rotating, the double black hole system nevertheless has a
generically non-vanishing total angular momentum
 \be
J = \kappa^2ma = \frac{2\kappa^2\nu}{m}\left[1+m^2 - \s\right].
 \ee
This has two origins. First, the opposite NUT charges $\pm N_H$ of the two horizons
generate a dipole angular momentum $2\kappa N_H$, with
 \be
N_H = \frac{\kappa\nu}{m^2}\left[-(m^2+5m+12) + (m+4)\s\right].
 \ee
This is always nonzero and opposite in sign to $\nu$. Second, the combined electric
and magnetic fields generate an electromagnetic angular momentum.
The two exactly balance, $J=0$, for the critical value
 \be
m_c = \left(\frac{-3+\sqrt{41}}{2}\right)^{1/2} \simeq 1.3044
 \ee
(as previously reported in \cite{dimagn}).
On the other hand, the magnetic moment $\mu=\kappa^2\m$, with
 \be
\m^2 = \frac8{m^2}\left[m^2 -12 + 4\s\right]
 \ee
never vanishes.
The table below (where we have assumed $\nu$ and $\m$ positive) gives
the values of the scaled dimensionless observables: mass,
angular momentum, and magnetic moment for $m=0$, the critical value $m=m_c$, and
the maximal scaled mass $m=\mm$ above which a ring singularity appears:
 \be
 \begin{tabular}{c|c|c|c}
$m$ & $0$ & $1.304$ & $2.506$ \\\hline
$ma$ & $-1.633$ & $0$ &$4.250$\\\hline
$\m$ & $1.633$ & $1.545$ & $1.057$
\end{tabular}
 \ee

Now we evaluate the various horizon characteristics. The area of each horizon component is
${\cal A}_H = \pi\kappa^2\Sigma_0$, with
 \be
\Sigma_0 = \frac8{m^2}\left[4m^2+6m-12 - (m^2+2m-4)\s\right]
 \ee
As shown in \cite{smarr}, the usual Smarr formula $M_H = 2\Omega_HJ_H + 2T_HS + \Phi_HQ_H$
(where $-\Phi_H = \hat{A}_t$ the horizon electric potential in the co-rotating frame)
remains valid in the dyonic case. In the present case, the two horizon components are
degenerate ($T_H=0$) and non-rotating ($\Omega_H=0$), so that the two Komar horizon masses
are simply given in terms of the horizon electric charges and potentials by
 \be
M_H = \Phi_HQ_H,
 \ee
with
 \ba
Q_H &=& \frac{4\kappa\m\nu}{m^2\Sigma_0}\left[-2(m+3)(m+6) + (m^2+6m+12)\s\right], \nn\\
\Phi_H &=& - \frac{2\m\nu}{m\Sigma_0}\left[m-1 + \s\right].
 \ea
The expression of $P_H$ in terms of $m$ is too complicated to give here.

The table below\footnote{We have again assumed $\nu$ and $\m$ positive. The electric and magnetic charges are odd
under inversion of $\m$, and the electric charge and the NUT charge are odd under inversion of $\nu$.}
gives the values of the horizon areas, Smarr masses, NUT charges, and electric
and magnetic charges for $m=0, m_c, m_{\rm max}$:
 \be
\begin{tabular}{c|c|c|c}
$m$ & $0$ & $1.304$ & $2.506$ \\\hline
${\cal A}_H/\pi\kappa^2$ & $2.667$ & $8.436$ &$20.55$\\\hline
$M_H/\kappa$ & $0$ & $0.4742$ & $0.1470$ \\\hline
$N_H/\kappa$ & $-0.8165$ & $-1.820$ & $-3.024$ \\\hline
$Q_H/\kappa$ & $0$ & $-1.069$ & $-1.200$ \\\hline
$P_H/\kappa$ & $0.8165$ & $0.8074$ & $0.4770$
\end{tabular}
 \ee
The sum of the two horizon Komar masses is generically different from the asymptotic mass,
the difference corresponding to the Komar mass $M_S=M-2M_H$ of the Dirac-Misner string.
The string has also an electric
charge $Q_S=-2Q_H$ balancing the horizon charges (more on this below).

Let us discuss in more detail the large distance limit of the solution. This is defined as the limit where the
distance $2\kappa$ between the two horizons becomes very large, $\kappa\to\infty$, while the total mass
$M$ is held fixed, $m = M/\kappa\to0$. In this limit,
 \be
\nu \to \frac{m}{\sqrt6}, \quad \delta \to 1, \quad mb \to -\sqrt\frac{8}{3}, \quad \Phi_H \to -1,
 \ee
leading to the simple relations
 \be
M_H \simeq - Q_H \simeq M, \quad N_H \simeq - P_H \simeq - \sqrt\frac{2}{3}\kappa, \quad {\cal A}_H \simeq
\frac{8}{3}\pi\kappa^2.
 \ee
The corresponding asymptotic observables are the total mass $M$ and the comparatively much larger,
approximately equal (in gravitational units and absolute value) angular and magnetic dipole moments:
 \be
\mu \simeq -J \simeq \sqrt\frac{8}{3}\kappa^2.
  \ee
The distance between the two constituent black holes, which scales as their areal radius, is of the order
of the ratio $|J|/M$. Note that both the gravimagnetic and magnetic moments are in this limit essentially
due to the dipole moments of the opposite charges carried by the two horizons, $J \simeq 2\kappa N_H$ and
$\mu \simeq 2\kappa P_H$. In the limit $\kappa\to\infty$ in which $m$ may
be neglected altogether, the solution reduces to a Majumdar-Papapetrou \cite{MP} superposition of
two extreme massless black holes separated by a distance $2\kappa$ and carrying opposite NUT and
 magnetic charges with the special value $N_H = - P_H = - \sqrt{2/3}\kappa$.

As we have seen, although the Dirac-Misner string between the two black holes is tensionless, the overall
balance nevertheless requires that it carry a negative Komar mass $M_S \simeq -M$ and an electric charge
$Q_S \simeq 2M$. Given that energy is actually non-localisable in general relativity, the attribution
of a Komar mass to the string does not seem relevant. On the other hand, electric charge
is a local quantity, so it would seem surprising that the Dirac-Misner string is charged. It turns out that
this apparent string charge
 \be\lb{QS}
Q_S = \frac1{4\pi}\int_{(\xi=0)}\sqrt{|g|}F^{t\xi}\,dy\,d\varphi
 \ee
($\xi^2\equiv x^2-1$) is induced on the Misner string by its interaction with the external magnetic field
generated by the two horizons. To see this, we evaluate (\ref{QS}) {\em \`a la} Tomimatsu \cite{tom84}.
The covariant density $\sqrt{|g|}F_{t\xi}$ vanishes on the string $\xi^2=0$ because both $\sqrt{|g|}$ from
(\ref{string}) and $F_{t\xi} = -\partial_\xi A_t$ go to zero as O$(\xi)$. It follows that, on the string
 \be
\sqrt{|g|}F^{t\xi} = \omega_S \sqrt{|g|}F^{\varphi\xi} = - 4N_HB_y,
 \ee
where we have identified $N_H= -\omega_S/4$, and introduced the magnetic vector field dual to the magnetic
tensor field. This vector field is the gradient of the magnetic scalar field $u$ (the imaginary part of
the Ernst potential $\psi$), leading eventually from (\ref{QS}) to
 \be
Q_S = - 2N_H u(x=1,y)\big]_{-1}^{1}.
 \ee
Evaluating this in the limit $m\to0$ leads to
 \be
Q_S \simeq - \frac{4N_H\m m}{m^2b^2} \simeq 2M.
 \ee

\setcounter{equation}{0}
\section{Conclusion}

We have presented a two-parameter family of exact asymptotically flat stationary solutions to
Einstein-Maxwell theory. These solutions are free from conical singularities, and free from
ring singularities in a specific parameter range.
They represent systems of two non-rotating extreme black holes with equal masses and electric charges,
and opposite magnetic and NUT charges, connected by a tensionless Dirac-Misner string. The asymptotic
observables are the total mass $M$, angular momentum $J$ (which can possibly vanish), and magnetic
moment $\mu$.

Let us emphasize that the balance achieved in these field configurations necessitates a fine tuning of
all four black hole charges: for a given ADM mass $M$ of the configuration, the values of the various
black hole charges vary with the distance $2\kappa$ between the two black holes. Furthermore, this
distance is bounded below by
 \be
2\kappa > \frac{2M}{m_{\rm max}} \simeq 0.8\,M.
 \ee
This differs from the case of
Majumdar-Papepetrou or Israel-Wilson-Perj\`es superpositions, where the charges
obey a no-force law which does not involve the distance between the two sources. The necessary
presence of a NUT dipole implies a Misner string connecting the two horizons and breaking the axis
condition. We have shown elsewhere that Misner strings are not an obstacle to geodesic motion, and
argued that they do not necessarily lead to observable violations of causality \cite{GC15,GC18}.

We have also discussed the large-distance limit. In this limit, the system has
approximately equal (in gravitational units and absolute value) angular momentum and magnetic moment,
which can be arbitrarily large relative to the mass. The distance between the two extreme constituent
black holes, and their magnetic/gravimagnetic charges, are of the order of the square root of the
total angular momentum, while the black hole masses are of the order of the total mass.

\section*{Acknowledgments}
I warmly thank Dmitry Gal'tsov for a critical reading of the manuscript and useful suggestions.

\end{document}